\documentclass[
aps,
prc,
reprint,
superscriptaddress,
nofootinbib,
showkeys
]{revtex4-2}

\usepackage[utf8]{inputenc}
\usepackage[T1]{fontenc}
\usepackage{amsmath,amssymb,bm}
\usepackage{graphicx}
\usepackage{xcolor}
\usepackage{booktabs}
\usepackage{tikz-feynman}
\tikzfeynmanset{compat=1.1.0}
\usepackage{array}
\usepackage{multirow}
\usepackage{slashed}
\usepackage{dcolumn}
\usepackage{mathrsfs}
\usepackage{microtype}

\definecolor{linkblue}{RGB}{0,70,150}
\definecolor{citered}{RGB}{150,35,35}
\definecolor{urlpurple}{RGB}{100,40,150}

\usepackage[
colorlinks=true,
linkcolor=linkblue,
citecolor=citered,
urlcolor=urlpurple
]{hyperref}

\IfFileExists{orcidlink.sty}{
	\usepackage{orcidlink}
}{
	\newcommand{\orcidlink}[1]{\href{https://orcid.org/#1}{\textsuperscript{\scriptsize ORCID}}}
}

\newcommand{\dd}{\mathrm{d}}

\newcommand{\Pperp}{\bm P_\perp}
\newcommand{\Dperp}{\bm \Delta_\perp}
\newcommand{\Baxis}{\bm B_{\rm ax}}
\newcommand{\rperp}{\bm r}
\newcommand{\bperp}{\bm b}
\newcommand{\Xperp}{\bm X}
\newcommand{\kappaperp}{\bm \kappa}

\newcommand{\Acal}{\mathcal A}
\newcommand{\Vcal}{\mathcal V}

\begin{document}
	
	% ============================================================
	% Title
	% ============================================================
\title{
	Orbital Angular Momentum as a Transverse Probe of Elliptic Small-$x$ Gluon Tomography
}
	
	% Alternative:
	% \title{Wave-Packet Orbital-Angular-Momentum Probes of Elliptic Small-$x$ Gluon Tomography}
	
	% ============================================================
	% Authors
	% ============================================================
	\author{Wei Kou\orcidlink{0000-0002-4152-2150}}
	\email{kouwei@impcas.ac.cn}
	\affiliation{Institute of Modern Physics, Chinese Academy of Sciences, Lanzhou 730000, Gansu Province, China}
	\affiliation{Southern Center for Nuclear Science Theory (SCNT), Institute of Modern Physics, Chinese Academy of Sciences, Huizhou 516000, Guangdong Province, China}
	\affiliation{School of Nuclear Science and Technology, University of Chinese Academy of Sciences, Beijing 100049, China}
	\affiliation{State Key Laboratory of Heavy Ion Science and Technology, Institute of Modern Physics, Chinese Academy of Sciences, Lanzhou 730000, China}
	
	\author{Xurong Chen}
	\email{xchen@impcas.ac.cn}
	\affiliation{Institute of Modern Physics, Chinese Academy of Sciences, Lanzhou 730000, Gansu Province, China}
	\affiliation{Southern Center for Nuclear Science Theory (SCNT), Institute of Modern Physics, Chinese Academy of Sciences, Huizhou 516000, Guangdong Province, China}
	\affiliation{School of Nuclear Science and Technology, University of Chinese Academy of Sciences, Beijing 100049, China}
	\affiliation{State Key Laboratory of Heavy Ion Science and Technology, Institute of Modern Physics, Chinese Academy of Sciences, Lanzhou 730000, China}
	
	% ============================================================
	% Abstract
	% ============================================================
	\begin{abstract}
		We propose a new way to probe elliptic small-$x$ gluon geometry in hard diffractive dijet deep inelastic scattering: a localized lepton wave packet carrying orbital angular momentum is used as a tunable transverse analyzer of the target. Unlike the conventional plane-wave setup, where the probe provides a fixed transverse projection, the OAM mode supplies additional radial and azimuthal structure while leaving the target dipole matrix element unchanged. We characterize the resulting elliptic response through $S_M=D_M-C_M$ and find finite-yield response nodes whose origin depends on the OAM channel. In the present benchmark, the $M=1$ crossing is associated mainly with a zero of the direct response, whereas the $M=3$ channel exhibits a nontrivial cancellation $D_3=C_3\neq0$. Signed transverse response densities show that this node results from a spatial cancellation rather than from a disappearance of the underlying scattering strength, and its position can be shifted by changing the beam--target offset. The specific sensitive value of $M$ depends on the analyzer profile and kinematics; the broader result is that localized OAM wave packets provide a controllable probe-side degree of freedom for small-$x$ gluon tomography.
	\end{abstract}

	\maketitle
	
	% ============================================================
	% Main text
	% ============================================================
	
	\section{Introduction} 
	\label{sec:introduction} 
	
	Hadron tomography aims to determine not only how much momentum is carried by partons, but also how that momentum is distributed in transverse space. Wigner distributions and generalized transverse-momentum-dependent distributions provide the natural language for this multidimensional structure ~\cite{Ji:2003ak,Belitsky:2003nz,Lorce:2011kd,Pasechnik:2024wigner}. At small $x$, where gluons dominate, transverse imaging is closely connected with high-density QCD dynamics, saturation, and coherent diffraction ~\cite{McLerran:1993ni,McLerran:1993ka,JalilianMarian:1997jx, Balitsky:1995ub,Kovchegov:1999yj}. Accessing this geometry is one of the central goals of the Electron-Ion Collider program ~\cite{Accardi:2012qut,AbdulKhalek:2021gbh}. 
	An important question is therefore whether one can obtain additional information not only by measuring new final states, but also by changing the transverse quantum structure of the probe itself. 
	
	Diffractive dijet production in deep inelastic scattering provides a particularly clean setting for this question. In $\gamma^\ast p\to q\bar q p'$, the relative dijet momentum $\Pperp$ and the target recoil $\Dperp$ resolve complementary transverse information. Their second azimuthal harmonic is sensitive to the elliptic component of the small-$x$ gluon Wigner distribution ~\cite{Hatta:2016dxp}. In the dipole picture, this signal originates from a correlation between the dipole orientation and its transverse position in the target. This connection has been developed in a variety of small-$x$ and saturation frameworks ~\cite{Zhou:2016rnt,Hagiwara:2017fye,Altinoluk:2015dpi, Mantysaari:2019csc,Salazar:2019ncp,Mantysaari:2019hkq, Hatta:2019ixj,Boer:2021upt,Iancu:2021rup,Hatta:2022lzj, Hatta:2024ocp,Shao:2024gri,Hatta:2017cte}. 
	In all of these plane-wave formulations, however, the transverse structure of the probe itself is fixed. The target geometry is sampled through the same plane-wave projection. 
	
	Orbital-angular-momentum wave packets offer a qualitatively different possibility. Vortex states carry a structured transverse phase and intensity profile, and such states are well established in optics and electron microscopy ~\cite{Allen:1992zz,Durnin:1987zz,Uchida:2010hbm, Verbeeck:2010ezk,Bliokh:2017uvr,Lloyd:2017dob,Jentschura:2010ap, Ivanov:2022jzh}. Their use in scattering, however, requires some care. For a single ideal Bessel state scattering from a completely delocalized plane-wave target, the OAM phase does not by itself produce a new phase-sensitive observable; the result reduces to an azimuthal average over plane-wave components ~\cite{Ivanov:2011bv,Ivanov:2011aa}. OAM sensitivity becomes observable only when a transverse reference is retained, for example through a localized target, a finite beam--target overlap, or an equivalent selection of the collision geometry ~\cite{Ivanov:2012na,Ivanov:2016oue,Karlovets:2015nva, Karlovets:2016jrd,Karlovets:2016uhb,Yang:2026byv, Karlovets:2022jym,Liu:2022nei}. Recent work on the generation and detection of high-energy vortex leptons further motivates exploring such possibilities in high-energy reactions ~\cite{Ababekri:2024lgx,Li:2024fek}. 
	
	The idea pursued here is to use this transverse structure as a new analyzer of small-$x$ gluon geometry. We replace the incoming plane-wave lepton by a localized OAM wave packet and ask how the same diffractive target amplitude is read out by different transverse modes. The target side is deliberately left unchanged: it is described by the same impact-parameter-dependent dipole matrix element used in ordinary diffractive tomography ~\cite{Kowalski:2003hm,Kowalski:2006hc,Rezaeian:2013tka, Salazar:2019ncp,Ji:2016jgn,Hatta:2016aoc}. The OAM dependence enters only through the lepton-side transverse analyzer. Thus the purpose is not to define a new gluon distribution, nor to regard the exchanged virtual photon as an asymptotic vortex particle. Rather, the localized OAM state changes the transverse projection through which an existing small-$x$ target matrix element is observed. 
	
	This additional probe-side degree of freedom leads to a simple physical question: can an OAM mode enhance, suppress, or even reverse the response to a given component of the target geometry? We address this question for the elliptic component of the dipole amplitude. For the OAM-resolved elliptic moment $A_2^{(M)}$, we introduce the linear response 
	\begin{equation*} 
		S_M=D_M-C_M , 
	\end{equation*} 
	where $D_M$ represents the direct elliptic response and $C_M$ the change induced through the normalization of the observable. A response node occurs when these two contributions balance while the baseline diffractive weight remains nonzero. Such a node is therefore not a disappearance of the target signal, but a zero of the particular transverse projection selected by the OAM analyzer. 
	
	Our numerical study shows that this projection is strongly mode-dependent. Within the benchmark considered here, sign changes appear in the $M=1$ and $M=3$ channels, while the other channels do not cross zero in the scanned range. The $M=1$ crossing is dominated by a near-zero of the direct response, whereas the $M=3$ channel exhibits a cancellation between two individually finite contributions. We emphasize that the special role of $M=3$ is not a universal OAM selection rule: the most sensitive channel can change with the radial profile, kinematics, and transverse geometry of the analyzer. What is more general is the existence of a tunable mode-dependent projection mechanism. Signed response-density maps show how positive and negative transverse regions cancel at the $M=3$ node, while varying the beam--target offset moves the node continuously. These observations illustrate how structured incoming states can add a new handle to small-$x$ tomography without changing the underlying target distribution. 
	
	The paper is organized as follows. Section~\ref{sec:formalism} introduces the localized transverse geometry and the OAM-resolved lepton-side analyzer. Section~\ref{sec:response} defines the target input and the OAM-resolved elliptic response. Section~\ref{sec:numerics} presents the channel dependence, the finite-yield response nodes, their transverse-space interpretation, and their dependence on the collinear beam--target offset. Section~\ref{sec:conclusions} summarizes the main results and their physical implications.

	\section{Plane-Wave Tomography and Localized OAM Readout}
	\label{sec:formalism}
	
	\subsection{Plane-wave diffractive dijet tomography}
	\label{subsec:plane-wave}
	
\begin{figure}[htbp]
		\centering
		\includegraphics[width=0.98\linewidth]{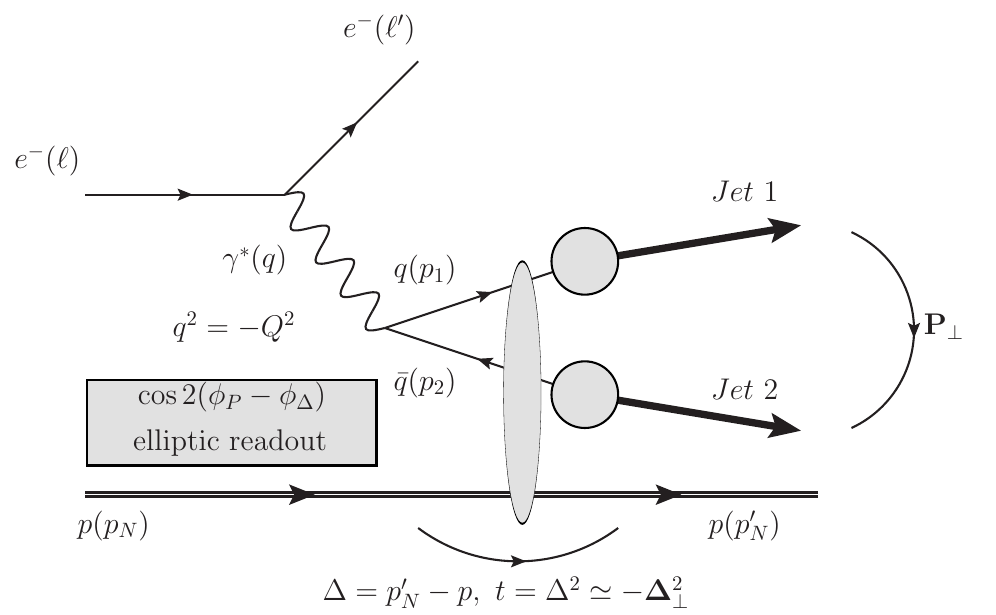}
		\caption{
			Plane-wave hard diffractive dijet production in deep inelastic scattering.
			The virtual photon splits into a $q\bar q$ dipole, which scatters coherently from the target and forms the diffractive dijet final state.
			The relative dijet momentum $\Pperp$ and the target recoil $\Dperp$ define the elliptic harmonic $\cos 2(\phi_P-\phi_\Delta)$.
		}
		\label{fig:plane-wave-dijet}
	\end{figure}

	We begin with the hard diffractive dijet process
	\begin{equation}
		e(\ell)+p(p_N)
		\to
		e(\ell')+q(p_1)+\bar q(p_2)+p(p_N'),
		\label{eq:ep-dijet-process}
	\end{equation}
	whose hadronic subprocess is
	\begin{equation}
		\gamma^\ast(q)+p(p_N)
		\to
		q(p_1)+\bar q(p_2)+p(p_N').
		\label{eq:gamma-p-dijet-process}
	\end{equation}
	A schematic representation is shown in Fig.~\ref{fig:plane-wave-dijet}. The incoming electron emits a virtual photon, which splits into a quark--antiquark dipole. The dipole scatters coherently from the target through a color-singlet exchange and produces a diffractive dijet final state.
	
	The virtual-photon momentum and virtuality are
	\begin{equation}
		q=\ell-\ell',
		\qquad
		Q^2=-q^2>0,
		\label{eq:photon-virtuality}
	\end{equation}
	while the target momentum transfer is
	\begin{equation}
		\Delta=p_N'-p_N,
		\qquad
		t=\Delta^2\simeq-\Dperp^2.
		\label{eq:target-recoil}
	\end{equation}
	The relative transverse momentum of the dijet is defined as
	\begin{equation}
		\Pperp
		=
		(1-z)\bm p_{1\perp}-
		z\bm p_{2\perp},
		\label{eq:dijet-relative-momentum}
	\end{equation}
	where $z$ is the longitudinal momentum fraction carried by the quark in the photon splitting. In the correlation limit, $|\Pperp|$ provides the hard dijet scale, whereas $|\Dperp|$ resolves the transverse position dependence of the target.
	
	The standard elliptic observable is the second harmonic in the relative azimuthal angle between $\Pperp$ and $\Dperp$,
	\begin{equation}
		A_2
		=
		\frac{
			\displaystyle
			\int_0^{2\pi}\dd\phi_P
			\cos 2(\phi_P-\phi_\Delta)
			\frac{\dd\sigma}{\dd\phi_P}
		}{
			\displaystyle
			\int_0^{2\pi}\dd\phi_P
			\frac{\dd\sigma}{\dd\phi_P}
		}.
		\label{eq:plane-wave-A2}
	\end{equation}
	In the small-$x$ correlation limit, this harmonic probes the elliptic component of the gluon Wigner distribution or gluon generalized transverse-momentum-dependent distribution~\cite{Hatta:2016dxp,Zhou:2016rnt,Hagiwara:2017fye,Mantysaari:2019csc,Hatta:2024ocp,Shao:2024gri}.
	
	In the dipole representation, the corresponding plane-wave amplitude can be written schematically as
	\begin{equation}
		\begin{aligned}
			\Acal_{\rm PW}(z,\Pperp,\Dperp)
			={}&
			\int
			\dd^2\rperp
			\dd^2\bperp
			e^{-i\Pperp\cdot\rperp}
			e^{-i\Dperp\cdot\bperp}
			\\
			&\times
			\Psi(z,\rperp;Q)
			N_Y(\rperp,\bperp),
		\end{aligned}
		\label{eq:plane-wave-amplitude}
	\end{equation}
	where $\Psi(z,\rperp;Q)$ denotes the light-front virtual-photon
	wave function and $N_Y(\rperp,\bperp)$ is the
	impact-parameter-dependent dipole amplitude at rapidity
	$Y=\ln(1/x_g)$
	~\cite{Nikolaev:1990ja,Mueller:1989st,Kowalski:2006hc,
		Dominguez:2011wm}.
	Spin, flavor, and photon-polarization labels are suppressed at this
	formal level. The numerical benchmark below specializes this
	expression to a single effective longitudinal-photon channel, whose
	radial wave function and mass parameter are specified explicitly in
	Sec.~\ref{subsec:benchmark-photon}.
	
	The phase $e^{-i\Pperp\cdot\rperp}$ makes $\Pperp$ conjugate to the dipole size and orientation, whereas $e^{-i\Dperp\cdot\bperp}$ makes $\Dperp$ conjugate to the dipole position in the target. The elliptic harmonic arises when $N_Y(\rperp,\bperp)$ contains a quadrupole correlation between $\rperp$ and $\bperp$. The OAM construction developed below retains this target matrix element and modifies only the external transverse projection acting on it.
	
	\subsection{Localized transverse reference geometry}
	\label{subsec:localized-geometry}
	
	A phase-sensitive OAM readout requires a transverse reference geometry. For a single ideal Bessel state incident on a completely delocalized plane-wave target, the OAM phase does not generate an independent differential structure; the cross section reduces to an azimuthal average over plane-wave components~\cite{Ivanov:2011bv,Ivanov:2011aa,Ivanov:2022jzh}. Retaining coherent OAM information therefore requires a localized transverse overlap, a target wave packet, or an equivalent postselection of the collision geometry~\cite{Ivanov:2012na,Ivanov:2016oue,Karlovets:2015nva,Karlovets:2016jrd,Karlovets:2016uhb,Yang:2026byv}.
	
\begin{figure*}[htbp]
		\centering
		\includegraphics[width=0.96\textwidth]{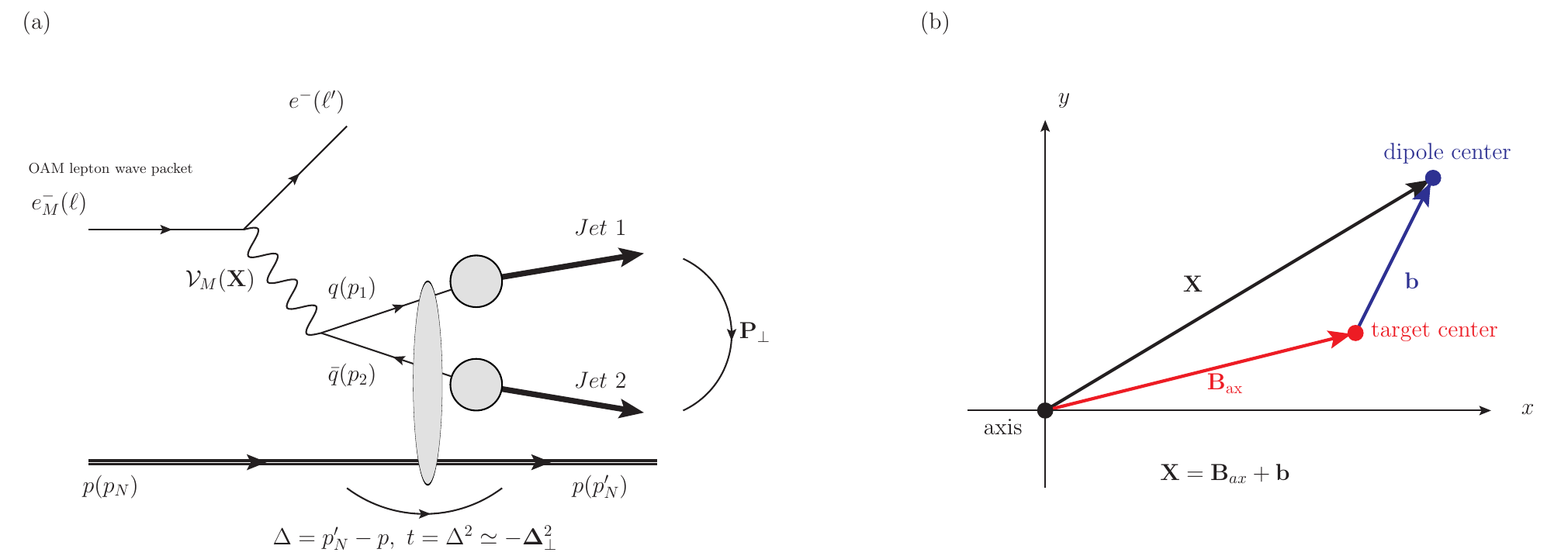}
		\caption{
			Localized OAM extension of diffractive dijet tomography.
			Panel (a) shows the standard diffractive subprocess probed by a localized lepton wave packet.
			Panel (b) defines the transverse geometry: $\Baxis$ is the displacement of the target center from the OAM axis, $\bperp$ is the dipole impact parameter measured from the target center, and $\Xperp=\Baxis+\bperp$ is the global dipole-center coordinate.
		}
		\label{fig:localized-oam-geometry}
	\end{figure*}

	The localized geometry is illustrated in Fig.~\ref{fig:localized-oam-geometry}. The target-side diffractive subprocess is unchanged from the plane-wave case. The new ingredient is the lepton-side transverse analyzer represented by the OAM-projected exchange kernel $\Vcal_M$. The OAM dependence therefore belongs to the probing current rather than to a new target distribution.
	
	We denote by $\Baxis$ the transverse displacement of the target center from the OAM axis of the incoming lepton wave packet. The internal dipole impact parameter $\bperp$ is measured from the target center. The corresponding global dipole-center coordinate is
	\begin{equation}
		\Xperp
		=
		\Baxis+\bperp.
		\label{eq:global-coordinate}
	\end{equation}
	This relation distinguishes the external beam--target alignment from the internal impact parameter entering the target matrix element.
	
	To specify the transverse coordinate sampled by the lepton-side
	kernel, we introduce the quark and antiquark transverse positions
	relative to the target center,
	\begin{equation}
		\bm x_q
		=
		\bperp+\frac{\rperp}{2},
		\qquad
		\bm x_{\bar q}
		=
		\bperp-\frac{\rperp}{2}.
		\label{eq:dipole-endpoint-coordinates}
	\end{equation}
	For a photon splitting in which the quark carries longitudinal
	fraction $z$, the transverse light-front center of the
	$q\bar q$ pair is
	\begin{equation}
		\begin{aligned}
			\bm x_\gamma
			&=
			z\,\bm x_q
			+
			(1-z)\bm x_{\bar q}
			\\
			&=
			\bperp
			-
			\left(
			\frac{1}{2}-z
			\right)
			\rperp.
		\end{aligned}
		\label{eq:photon-transverse-center}
	\end{equation}
	The coordinate of this point relative to the OAM axis is therefore
	\begin{equation}
		\Xperp_{\rm ker}
		=
		\Baxis
		+
		\bperp
		-
		\left(
		\frac{1}{2}-z
		\right)
		\rperp.
		\label{eq:kernel-coordinate-general}
	\end{equation}
	This is the coordinate sampled by the localized transverse analyzer
	throughout the numerical study. It reduces to the dipole-center
	coordinate $\Baxis+\bperp$ for symmetric splitting, $z=1/2$, while
	for asymmetric configurations the analyzer samples the corresponding
	light-front transverse center of the $q\bar q$ state.
	
	At fixed external alignment, the OAM-projected diffractive amplitude is written as
	\begin{equation}
		\begin{aligned}
			\Acal_M
			(z,\Pperp,\Dperp;\Baxis)
			={}&
			\int
			\dd^2\rperp
			\dd^2\bperp
			\Vcal_M(\Xperp_{\rm ker})
			e^{-i\Pperp\cdot\rperp}
			e^{-i\Dperp\cdot\bperp}
			\\
			&\times
			\Psi(z,\rperp;Q)
			N_Y(\rperp,\bperp).
		\end{aligned}
		\label{eq:fixed-offset-amplitude}
	\end{equation}
	Relative to Eq.~\eqref{eq:plane-wave-amplitude}, the target matrix element is unchanged. The localized OAM information enters entirely through the external kernel $\Vcal_M(\Xperp_{\rm ker})$.
	
	For an ensemble with a finite distribution of target positions relative to the OAM axis, the physical yield is obtained through an incoherent average,
	\begin{equation}
		\frac{\dd\sigma_M[P_B]}{\dd\phi_P}
		=
		\int
		\dd^2\Baxis
		P_B(\Baxis)
		\frac{\dd\sigma_M(\Baxis)}{\dd\phi_P},
		\label{eq:PB-average}
	\end{equation}
	where $P_B(\Baxis)$ is normalized to unity. A convenient localization model is
	\begin{equation}
		P_B(\Baxis)
		=
		\frac{1}{2\pi\sigma_B^2}
		\exp\left(
		-\frac{B_{\rm ax}^2}{2\sigma_B^2}
		\right).
		\label{eq:PB-gaussian}
	\end{equation}
	The limit $\sigma_B\to0$ selects a sharply aligned configuration, whereas increasing $\sigma_B$ progressively removes the phase coherence associated with a definite transverse reference geometry.
	
	For the dipole-center kernel, the loss of coherence can be displayed directly in momentum space. Writing
	\begin{equation}
		\Vcal_M(\Xperp)
		=
		\int
		\frac{\dd^2\kappaperp}{(2\pi)^2}
		\widetilde{\Vcal}_M(\kappaperp)
		e^{i\kappaperp\cdot\Xperp},
		\label{eq:kernel-fourier-general}
	\end{equation}
	the fixed-offset amplitude becomes
	\begin{equation}
		\Acal_M(\Pperp,\Dperp;\Baxis)
		=
		\int
		\frac{\dd^2\kappaperp}{(2\pi)^2}
		\widetilde{\Vcal}_M(\kappaperp)
		e^{i\kappaperp\cdot\Baxis}
		\Acal_{\rm PW}
		(\Pperp,\Dperp-\kappaperp).
		\label{eq:oam-convolution}
	\end{equation}
	The localized readout is therefore a coherent convolution of plane-wave diffractive amplitudes evaluated at shifted recoil momenta.
	
	After averaging over the offset distribution, interference between transverse components $\kappaperp$ and $\kappaperp'$ is weighted by
	\begin{equation}
		\widetilde P_B(\bm q)
		=
		\int
		\dd^2\Baxis
		P_B(\Baxis)
		e^{i\bm q\cdot\Baxis}.
		\label{eq:PB-fourier}
	\end{equation}
	For the Gaussian profile in Eq.~\eqref{eq:PB-gaussian}, this factor becomes
	\begin{equation}
		\widetilde P_B(\bm q)
		=
		\exp\left(
		-\frac{\sigma_B^2q^2}{2}
		\right).
		\label{eq:PB-fourier-gaussian}
	\end{equation}
	Large $\sigma_B$ suppresses off-diagonal interference and leaves an incoherent average over plane-wave components. The localized fixed-offset regime and the conventional single-twisted azimuthal-averaging limit are therefore complementary limits of the same transverse-overlap description.
	
	\subsection{OAM-projected lepton current}
	\label{subsec:oam-current}
	
	The transverse analyzer in Eq.~\eqref{eq:fixed-offset-amplitude} originates from the lepton electromagnetic current. For an incoming plane-wave component with momentum $k$ and an outgoing component with momentum $k'$, the current is
	\begin{equation}
		J^\mu(k,k')
		=
		\bar u(k')\gamma^\mu u(k),
		\label{eq:plane-wave-current}
	\end{equation}
	with momentum transfer
	\begin{equation}
		q(k)=k-k'.
		\label{eq:component-transfer}
	\end{equation}
	Since the external spinors are on shell, each component satisfies
	\begin{equation}
		\begin{aligned}
			q_\mu(k)J^\mu(k,k')
			&=
			\bar u(k')
			\left(
			\slashed{k}-\slashed{k}'
			\right)
			u(k)
			=
			0.
		\end{aligned}
		\label{eq:ward-identity-component}
	\end{equation}
	The OAM-projected current is a coherent superposition of these conserved plane-wave currents.
	
	An incoming lepton wave packet with OAM index $M$ can be represented schematically as
	\begin{equation}
		|e_M\rangle=
		\int
		\frac{\dd^2\bm k_\perp}{(2\pi)^2}
		a_M(\bm k_\perp)
		e^{iM\phi_k}
		|e(k)\rangle,
		\label{eq:oam-lepton-state}
	\end{equation}
	where $a_M(\bm k_\perp)$ specifies the transverse momentum profile.
	The exact wave-packet current is obtained by coherently superposing
	the conserved plane-wave components in
	Eq.~\eqref{eq:plane-wave-current}; current conservation therefore
	holds at the level of the underlying component-wise construction.
	
	For the present benchmark, we do not evaluate the full
	component-dependent lepton tensor explicitly. Instead, the hard DIS
	variables are fixed at their central values, while the transverse
	mode dependence of the localized lepton state is retained through
	an effective analyzer kernel $\Vcal_M(\Xperp_{\rm ker})$ multiplying
	the standard dipole amplitude. The purpose of this reduction is to
	isolate the transverse OAM projection while leaving the target
	matrix element unchanged.
	
	The numerical analyzer used in this work is the Bessel--Gaussian kernel
	\begin{equation}
		\Vcal_M^{\rm BG}(\Xperp)
		=
		e^{iM\phi_X}
		J_M(q_TX)
		\exp\left(
		-\frac{X^2}{2R_\gamma^2}
		\right),
		\label{eq:BG-kernel}
	\end{equation}
	where
	\begin{equation}
		X=|\Xperp|,
		\qquad
		\phi_X=\arg(\Xperp).
		\label{eq:X-polar-def}
	\end{equation}
	We use Eq.~\eqref{eq:BG-kernel} without an additional
	mode-dependent overall normalization. Such a factor would multiply
	$\Acal_{0,M}$ and $\Acal_{1,M}$ equally and therefore cancel from
	$v_{2,0}^{(M)}$, $D_M$, $C_M$, and $S_M$, leaving the position and
	mechanism of the response nodes unchanged. The quantities
	$\Sigma_{i,M}$ and $W_{i,M}$ are consequently quoted as baseline
	weights in this fixed kernel convention rather than as absolutely
	normalized experimental cross sections
	~\cite{Ivanov:2011aa,Karlovets:2015nva,Karlovets:2016uhb,
		Bliokh:2017uvr,Lloyd:2017dob}.
	
	The integer $M$ labels the azimuthal OAM harmonic, $q_T$ controls the
	radial Bessel scale, and $R_\gamma$ sets the transverse envelope
	~\cite{Durnin:1987zz,Allen:1992zz}. The parameter $q_T$ is a mode
	scale of the external analyzer and is distinct from the photon
	virtuality, the target recoil, and the measured dijet relative
	momentum. 
	
	A broader class of localized OAM analyzers may be represented by a Laguerre--Gaussian-type profile,
	\begin{equation}
		\Vcal_M^{\rm LG}(\Xperp)
		=
		\left(
		\frac{X}{\sigma_\perp}
		\right)^{|M|}
		e^{iM\phi_X}
		\exp\left(
		-\frac{X^2}{2\sigma_\perp^2}
		\right),
		\label{eq:LG-kernel}
	\end{equation}
	where $\sigma_\perp$ is the transverse wave-packet width. As in
	Eq.~\eqref{eq:BG-kernel}, an overall profile normalization is left
	implicit because it does not affect the normalized response
	observables considered here. Equation~\eqref{eq:LG-kernel}
	illustrates that the construction is not conceptually restricted to
	the Bessel--Gaussian radial basis. The numerical results below use
	Eq.~\eqref{eq:BG-kernel}; systematic comparisons among different
	localized radial profiles are left for future work.
	
	The standard plane-wave readout is recovered only in the common limit
	\begin{equation}
		M=0,
		\qquad
		q_T\to0,
		\qquad
		R_\gamma\to\infty.
		\label{eq:plane-wave-common-limit}
	\end{equation}
	In this limit, the Bessel--Gaussian kernel becomes a constant transverse weight, and Eq.~\eqref{eq:fixed-offset-amplitude} reduces, up to an overall normalization, to Eq.~\eqref{eq:plane-wave-amplitude}. For finite $q_T$ or finite $R_\gamma$, the $M=0$ channel is still a structured wave-packet projection and should not be identified with the plane-wave limit.
	
	The phrase ``OAM-projected virtual-photon exchange'' will therefore refer to the transverse mode structure induced by the lepton current. The exchanged photon is not an asymptotic vortex particle. All OAM dependence in the present framework is encoded in the external lepton wave packet, its transverse overlap with the target, and the kernel $\Vcal_M$ acting on the standard diffractive dipole amplitude.

	\section{OAM-Resolved Elliptic Response}
	\label{sec:response}
	
	\subsection{Impact-parameter-dependent dipole input}
	\label{subsec:dipole-input}
	
	The target dynamics is encoded in the impact-parameter-dependent dipole amplitude
	\begin{equation}
		N_Y(\rperp,\bperp)
		=
		1-S_Y(\rperp,\bperp),
		\label{eq:dipole-amplitude-def}
	\end{equation}
	where $\rperp$ is the transverse separation of the $q\bar q$ dipole and $\bperp$ is the dipole impact parameter measured from the target center. The corresponding Wilson-line correlator is
	\begin{equation}
		S_Y(\rperp,\bperp)
		=
		\frac{1}{N_c}
		\left\langle
		{\rm Tr}
		U\left(
		\bperp+\frac{\rperp}{2}
		\right)
		U^\dagger\left(
		\bperp-\frac{\rperp}{2}
		\right)
		\right\rangle_Y .
		\label{eq:wilson-line-dipole}
	\end{equation}
	This is the same target matrix element that enters the plane-wave amplitude in Eq.~\eqref{eq:plane-wave-amplitude}. The localized OAM construction modifies the external transverse projection acting on this matrix element, but not the target distribution itself.
	
	For the numerical benchmark, we employ an impact-parameter-dependent saturation ansatz inspired by the GBW and impact-parameter dipole models~\cite{GolecBiernat:1998js,Kowalski:2003hm,Kowalski:2006hc,Rezaeian:2013tka,Salazar:2019ncp}. The isotropic component is written as
	\begin{equation}
		Q_s^2(b,Y)
		=
		Q_{s0}^2
		\exp(\lambda_s Y)
		\exp\left(
		-\frac{b^2}{B_p}
		\right),
		\label{eq:saturation-scale}
	\end{equation}
	and
	\begin{equation}
		N_0(r,b,Y)
		=
		1-
		\exp\left[
		-\frac{1}{4}
		\left(
		r^2Q_s^2(b,Y)
		\right)^\gamma
		\right],
		\label{eq:isotropic-dipole-input}
	\end{equation}
	where
	\begin{equation}
		r=|\rperp|,
		\qquad
		b=|\bperp|.
	\end{equation}
	The parameter $Q_{s0}$ fixes the initial saturation scale, $\lambda_s$ controls the rapidity dependence, $B_p$ sets the transverse width of the target profile, and $\gamma$ is an anomalous-dimension parameter. This ansatz is not used as a global fit to diffractive data. It provides a controlled background in which the target profile and the external OAM analyzer can be varied independently.
	
	The elliptic target component is introduced through a small quadrupole deformation,
	\begin{equation}
		\begin{aligned}
			N_Y(\rperp,\bperp)
			&={}
			N_0(r,b,Y)
			\\
			&+
			\epsilon_2
			N_\epsilon(r,b,Y)
			\cos 2(\phi_r-\phi_b)
			+
			\mathcal O(\epsilon_2^2),
		\end{aligned}
		\label{eq:elliptic-dipole-expansion}
	\end{equation}
	where $\phi_r$ and $\phi_b$ are the azimuthal angles of $\rperp$ and $\bperp$, respectively. The radial profile of the deformation is chosen as
	\begin{equation}
		N_\epsilon(r,b,Y)
		=
		2F_2(r,b)
		N_0(r,b,Y)
		\left[
		1-N_0(r,b,Y)
		\right],
		\label{eq:Nepsilon-def}
	\end{equation}
	with
	\begin{equation}
		F_2(r,b)
		=
		\frac{r^2b^2}
		{\left(
			r^2+b^2+B_\epsilon^2
			\right)^2}.
		\label{eq:F2-def}
	\end{equation}
	The factor $F_2(r,b)$ suppresses the deformation for a pointlike dipole and at the target center, where a preferred relative orientation cannot be resolved. The additional factor $N_0(1-N_0)$ localizes the deformation near the transition between the dilute and saturated regimes.
	
	For the numerical benchmark, the parameters of
	Eqs.~\eqref{eq:saturation-scale}--\eqref{eq:F2-def} are fixed to
	\begin{equation}
		\begin{aligned}
			Q_{s0}^2 &= 0.40~{\rm GeV}^2,
			&
			\lambda_s &= 0.25,
			\\
			B_p &= 5.0~{\rm GeV}^{-2},
			&
			\gamma &= 1,
			\\
			B_\epsilon &= 1.0~{\rm GeV}^{-1}.
		\end{aligned}
		\label{eq:dipole-benchmark-parameters}
	\end{equation}
	These values define an illustrative impact-parameter-dependent
	saturation background inspired by the GBW and impact-parameter
	dipole constructions
	~\cite{GolecBiernat:1998js,Kowalski:2003hm,
		Kowalski:2006hc,Rezaeian:2013tka}.
	They are kept fixed in every OAM channel and throughout all
	$q_T$ and alignment scans. The numerical study therefore tests the
	projection mechanism for one controlled target profile rather than
	refitting the target separately for each analyzer configuration.
	
	The parameter $\epsilon_2$ is introduced only as a bookkeeping parameter for the linear response to the elliptic target component. It is not interpreted as an independently tunable experimental variable. All dependence on the OAM index $M$, the radial mode scale $q_T$, the envelope width $R_\gamma$, and the beam--target alignment arises from the external analyzer and its transverse overlap with the fixed target distribution.
	
	\subsection{Longitudinal photon kernel used in the benchmark}
	\label{subsec:benchmark-photon}
	
	The formal amplitudes in Sec.~\ref{subsec:plane-wave} allow for the
	usual sums over photon polarization, quark flavor, and quark
	helicities. For the proof-of-principle numerical study, however, we
	retain a single effective longitudinal-photon channel. Its radial
	light-front kernel is taken as
	\begin{equation}
		\Psi_L(z,r;Q)
		=
		2Qz(1-z)
		K_0(\varepsilon_f r),
		\label{eq:longitudinal-photon-kernel}
	\end{equation}
	where
	\begin{equation}
		\varepsilon_f^2
		=
		z(1-z)Q^2+m_f^2,
		\label{eq:epsilon-f-def}
	\end{equation}
	and
	\begin{equation}
		m_f=0.14~{\rm GeV}.
		\label{eq:effective-quark-mass}
	\end{equation}
	Here $K_0$ is the modified Bessel function of the second kind. This
	is the standard radial structure of the longitudinal virtual-photon
	wave function in the dipole representation
	~\cite{Nikolaev:1990ja,Mueller:1989st,Kowalski:2006hc}.
	
	The parameter $m_f$ is used as a common effective light-quark mass.
	Within this single effective longitudinal channel, we omit the common
	flavor, color, helicity, electromagnetic, and phase-space prefactors. These omitted factors multiply the isotropic and elliptic
	amplitudes in the same way and therefore cancel from the normalized
	quantities $v_{2,0}^{(M)}$, $D_M$, $C_M$, and $S_M$. The calculated
	$\Sigma_{i,M}$ and $W_{i,M}$ should accordingly be interpreted as
	model weights in common arbitrary units rather than as absolutely
	normalized DIS cross sections.
	
	Restricting the calculation to the longitudinal channel isolates the
	transverse OAM-projection mechanism without introducing an additional
	polarization decomposition. A full phenomenological analysis would
	add the transverse-photon wave functions, explicit flavor sums, and
	the corresponding electroweak normalization, but these extensions
	are not required for identifying the response nodes studied here.
	
	\subsection{Normalized linear response}
	\label{subsec:linear-response}
	
	At fixed beam--target offset, the OAM-projected amplitude in Eq.~\eqref{eq:fixed-offset-amplitude} can be expanded in the target deformation as
	\begin{equation}
		\Acal_M
		=
		\Acal_{0,M}
		+
		\epsilon_2\Acal_{1,M}
		+
		\mathcal O(\epsilon_2^2).
		\label{eq:amplitude-expansion}
	\end{equation}
	The isotropic amplitude is
	\begin{equation}
		\begin{aligned}
			\Acal_{0,M}
			(z,\Pperp,\Dperp;\Baxis)
			&={}
			\int
			\dd^2\rperp
			\dd^2\bperp
			\Vcal_M(\Xperp_{\rm ker})
			e^{-i\Pperp\cdot\rperp}
			e^{-i\Dperp\cdot\bperp}
			\\
			&\times
			\Psi(z,\rperp;Q)
			N_0(r,b,Y),
		\end{aligned}
		\label{eq:A0M-def}
	\end{equation}
	whereas the first-order elliptic amplitude is
	\begin{equation}
		\begin{aligned}
			\Acal_{1,M}
			(z,\Pperp,\Dperp;\Baxis)
			&={}
			\int
			\dd^2\rperp
			\dd^2\bperp
			\Vcal_M(\Xperp_{\rm ker})
			e^{-i\Pperp\cdot\rperp}
			e^{-i\Dperp\cdot\bperp}
			\\
			&\times
			\Psi(z,\rperp;Q)
			N_\epsilon(r,b,Y)
			\cos 2(\phi_r-\phi_b).
		\end{aligned}
		\label{eq:A1M-def}
	\end{equation}
	The kernel coordinate $\Xperp_{\rm ker}$ is defined in Eq.~\eqref{eq:kernel-coordinate-general}. Spin, flavor, and photon-polarization indices are suppressed in Eqs.~\eqref{eq:A0M-def} and \eqref{eq:A1M-def}.
	
	The angular distribution has the corresponding expansion
	\begin{equation}
		\frac{\dd\sigma_M}{\dd\phi_P}
		=
		\frac{\dd\sigma_{0,M}}{\dd\phi_P}
		+
		\epsilon_2
		\frac{\dd\sigma_{1,M}}{\dd\phi_P}
		+
		\mathcal O(\epsilon_2^2),
		\label{eq:dsigma-expansion}
	\end{equation}
	where
	\begin{equation}
		\frac{\dd\sigma_{0,M}}{\dd\phi_P}
		=
		\sum_{\lambda,h,\bar h,f}
		\int_0^1
		\dd z
		\left|
		\Acal_{0,M}^{\lambda h\bar h f}
		(z,\Pperp,\Dperp)
		\right|^2,
		\label{eq:dsigma0-def}
	\end{equation}
	and
	\begin{equation}
		\begin{aligned}
			\frac{\dd\sigma_{1,M}}{\dd\phi_P}
			={}&
			\sum_{\lambda,h,\bar h,f}
			\int_0^1
			\dd z
			2{\rm Re}
			\left[
			\left(
			\Acal_{0,M}^{\lambda h\bar h f}
			\right)^\ast
			\Acal_{1,M}^{\lambda h\bar h f}
			\right].
		\end{aligned}
		\label{eq:dsigma1-def}
	\end{equation}
	The modulus squared and the interference term are evaluated at fixed
	$z$ before the longitudinal integration. Thus, the $z$ integration
	is performed incoherently at the cross-section level rather than by
	first constructing an amplitude integrated over $z$. If a finite
	offset distribution is used, Eqs.~\eqref{eq:dsigma0-def} and
	\eqref{eq:dsigma1-def} are understood after the incoherent average
	in Eq.~\eqref{eq:PB-average}.
	
	Equations~\eqref{eq:dsigma0-def} and \eqref{eq:dsigma1-def} display
	the general polarization, helicity, and flavor structure. In the
	numerical benchmark, these sums are replaced by the single effective
	longitudinal channel defined in
	Eqs.~\eqref{eq:longitudinal-photon-kernel}--\eqref{eq:effective-quark-mass}.
	We denote by $\Acal_{i,M}^{L}$, with $i=0,1$, the amplitudes in
	Eqs.~\eqref{eq:A0M-def} and \eqref{eq:A1M-def} evaluated with the
	replacement
	\begin{equation}
		\Psi(z,\rperp;Q)
		\longrightarrow
		\Psi_L(z,r;Q).
		\label{eq:longitudinal-amplitude-replacement}
	\end{equation}
	Accordingly, the angular distributions used in the numerical
	calculation are
	\begin{equation}
		\frac{\dd\sigma_{0,M}^{L}}{\dd\phi_P}
		=
		\int_0^1
		\dd z\,
		\left|
		\Acal_{0,M}^{L}
		(z,\Pperp,\Dperp;\Baxis)
		\right|^2,
		\label{eq:benchmark-dsigma0}
	\end{equation}
	and
	\begin{equation}
		\begin{aligned}
			\frac{\dd\sigma_{1,M}^{L}}{\dd\phi_P}
			={}&
			\int_0^1
			\dd z\,
			2\,{\rm Re}
			\left[
			\left(
			\Acal_{0,M}^{L}
			(z,\Pperp,\Dperp;\Baxis)
			\right)^\ast
			\right.
			\\
			&\left.
			\hspace{2.0em}\times
			\Acal_{1,M}^{L}
			(z,\Pperp,\Dperp;\Baxis)
			\right].
		\end{aligned}
		\label{eq:benchmark-dsigma1}
	\end{equation}
	In the remainder of the numerical analysis, the superscript $L$ is
	suppressed for notational simplicity. Thus,
	$\dd\sigma_{i,M}/\dd\phi_P$, $\Sigma_{i,M}$, and $W_{i,M}$ refer to
	the effective longitudinal-channel quantities unless stated
	otherwise.
	
	The OAM-resolved elliptic moment is defined as
	\begin{equation}
		A_2^{(M)}
		=
		\frac{
			\displaystyle
			\int_0^{2\pi}
			\dd\phi_P
			\cos 2(\phi_P-\phi_\Delta)
			\frac{\dd\sigma_M}{\dd\phi_P}
		}{
			\displaystyle
			\int_0^{2\pi}
			\dd\phi_P
			\frac{\dd\sigma_M}{\dd\phi_P}
		}.
		\label{eq:A2M-def}
	\end{equation}
	We introduce the integrated yield moments
	\begin{equation}
		\Sigma_{i,M}
		=
		\int_0^{2\pi}
		\dd\phi_P
		\frac{\dd\sigma_{i,M}}{\dd\phi_P},
		\qquad
		i=0,1,
		\label{eq:Sigma-iM-def}
	\end{equation}
	and the corresponding elliptic numerator moments
	\begin{equation}
		W_{i,M}
		=
		\int_0^{2\pi}
		\dd\phi_P
		\cos 2(\phi_P-\phi_\Delta)
		\frac{\dd\sigma_{i,M}}{\dd\phi_P},
		\qquad
		i=0,1.
		\label{eq:W-iM-def}
	\end{equation}
	Substitution into Eq.~\eqref{eq:A2M-def} gives
	\begin{equation}
		A_2^{(M)}
		=
		\frac{
			W_{0,M}+\epsilon_2W_{1,M}
		}{
			\Sigma_{0,M}+\epsilon_2\Sigma_{1,M}
		}
		+
		\mathcal O(\epsilon_2^2).
		\label{eq:A2M-ratio}
	\end{equation}
	Expanding this ratio to first order yields
	\begin{equation}
		\begin{aligned}
			A_2^{(M)}
			&=
			\frac{W_{0,M}}{\Sigma_{0,M}}
			\\
			&+
			\epsilon_2
			\left[
			\frac{W_{1,M}}{\Sigma_{0,M}}
			-
			\frac{
				W_{0,M}\Sigma_{1,M}
			}{
				\Sigma_{0,M}^2
			}
			\right]
			+
			\mathcal O(\epsilon_2^2).
		\end{aligned}
		\label{eq:A2M-expansion}
	\end{equation}
	We therefore write
	\begin{equation}
		A_2^{(M)}
		=
		v_{2,0}^{(M)}
		+
		\epsilon_2 S_M
		+
		\mathcal O(\epsilon_2^2),
		\label{eq:A2M-linear}
	\end{equation}
	where the baseline elliptic moment is
	\begin{equation}
		v_{2,0}^{(M)}
		=
		\frac{W_{0,M}}{\Sigma_{0,M}},
		\label{eq:v20M-def}
	\end{equation}
	and the normalized linear response is
	\begin{equation}
		S_M
		=
		\frac{W_{1,M}}{\Sigma_{0,M}}
		-
		v_{2,0}^{(M)}
		\frac{\Sigma_{1,M}}{\Sigma_{0,M}}.
		\label{eq:SM-def}
	\end{equation}
	Equivalently,
	\begin{equation}
		S_M
		=
		\left.
		\frac{\partial A_2^{(M)}}{\partial\epsilon_2}
		\right|_{\epsilon_2=0}.
		\label{eq:SM-derivative}
	\end{equation}
	
	The first term in Eq.~\eqref{eq:SM-def} is the direct response of the elliptic numerator to the target quadrupole deformation. The second term is generated by the simultaneous deformation of the total yield appearing in the denominator of $A_2^{(M)}$. Because $A_2^{(M)}$ is a normalized ratio, this subtraction contributes at the same order as the direct numerator response.
	
	It is convenient to define
	\begin{equation}
		D_M
		=
		\frac{W_{1,M}}{\Sigma_{0,M}},
		\label{eq:DM-def}
	\end{equation}
	and
	\begin{equation}
		C_M
		=
		v_{2,0}^{(M)}
		\frac{\Sigma_{1,M}}{\Sigma_{0,M}},
		\label{eq:CM-def}
	\end{equation}
	so that
	\begin{equation}
		S_M
		=
		D_M-C_M.
		\label{eq:SM-DM-CM}
	\end{equation}
	The OAM analyzer can therefore change the sign and magnitude of the normalized response either through the direct contribution $D_M$, through the normalization contribution $C_M$, or through their competition.
	
	\subsection{Finite-yield projection nodes}
	\label{subsec:projection-zero}
	
	A finite-yield projection node is defined by
	\begin{equation}
		S_M=0,
		\qquad
		\Sigma_{0,M}\neq0.
		\label{eq:finite-yield-zero-def}
	\end{equation}
	For a Bessel--Gaussian analyzer, the response can depend on $M$, $q_T$,
	$R_\gamma$, and the transverse beam--target geometry. For a general
	fixed-offset configuration, this dependence may be written as
	\begin{equation}
		S_M
		=
		S_M
		\left(
		q_T,R_\gamma;
		B_{\rm ax},
		\phi_B-\phi_\Delta
		\right).
		\label{eq:SM-parameter-dependence}
	\end{equation}
	The numerical results below specialize to the collinear configuration
	$\phi_B-\phi_\Delta=0$. For an offset-averaged observable, the fixed
	offset is replaced by the parameters of the distribution $P_B$.
	
	Using Eq.~\eqref{eq:SM-DM-CM}, the node condition becomes
	\begin{equation}
		D_M=C_M,
		\qquad
		\Sigma_{0,M}\neq0,
		\label{eq:finite-yield-zero-balance}
	\end{equation}
	or equivalently
	\begin{equation}
		W_{1,M}
		=
		v_{2,0}^{(M)}
		\Sigma_{1,M},
		\qquad
		\Sigma_{0,M}\neq0.
		\label{eq:finite-yield-zero-moments}
	\end{equation}
	This condition is a zero of the normalized response and not necessarily a zero of any individual amplitude or cross-section moment.
	
	Two limiting mechanisms are useful for interpreting such nodes. A direct-response node occurs when
	\begin{equation}
		D_M\simeq0,
		\qquad
		C_M\simeq0,
		\label{eq:direct-response-node}
	\end{equation}
	and the sign change of $S_M$ is governed predominantly by the zero of $D_M$. In this case, the equality $D_M=C_M$ is satisfied because both contributions are small.
	
	A compensation node instead satisfies
	\begin{equation}
		D_M=C_M\neq0.
		\label{eq:compensation-node}
	\end{equation}
	Here the direct response and the normalization response remain individually finite and cancel only after the normalized observable is formed. This is the genuinely nontrivial realization of a normalization-induced projection node.
	
	The distinction between these mechanisms cannot be inferred from $S_M$ alone. It requires the separate evaluation of $D_M$ and $C_M$. The numerical results below show that the $M=1$ and $M=3$ channels realize these two patterns approximately: the $M=1$ crossing is dominated by a zero of the direct response, whereas the $M=3$ node is generated by a cancellation between two finite contributions.
	
	A zero of $S_M$ does not imply
	\begin{equation}
		\frac{\dd\sigma_M}{\dd\phi_P}=0,
	\end{equation}
	nor does it imply that the target elliptic component $N_\epsilon$ disappears. It also does not require the full elliptic moment $A_2^{(M)}$ to vanish, because the baseline term $v_{2,0}^{(M)}$ can remain finite. Instead,
	\begin{equation}
		A_2^{(M)}(\epsilon_2)
		-
		A_2^{(M)}(0)
		=
		\epsilon_2 S_M
		+
		\mathcal O(\epsilon_2^2),
		\label{eq:delta-A2-linear}
	\end{equation}
	so that the leading variation of the normalized elliptic moment is suppressed at the projection node.
	
	The node position is not universal. It depends on the OAM channel, the radial structure and transverse width of the analyzer, the beam--target alignment, and the target profile. Changing $q_T$ modifies the radial Bessel weighting, whereas changing $\Baxis$ shifts the target relative to the OAM phase and radial structure. The nodes identified below should therefore be interpreted as analyzer-dependent projection structures of a fixed small-$x$ target matrix element, rather than as universal zeros of the target amplitude.

	\section{Numerical Results: OAM-Channel Selectivity and Finite-Yield Nodes}
	\label{sec:numerics}
	
	\subsection{Benchmark setup}
	\label{subsec:benchmark-setup}
	
	We now evaluate the OAM-resolved response in a localized fixed-offset geometry. The numerical study is designed as a proof-of-principle benchmark for isolating the analyzer-induced structure of the normalized elliptic response, rather than as a global phenomenological fit. The target input is the impact-parameter-dependent dipole amplitude defined in Eqs.~\eqref{eq:saturation-scale}--\eqref{eq:F2-def}, and the external transverse projection is implemented with the Bessel--Gaussian kernel in Eq.~\eqref{eq:BG-kernel}.
	
	Unless otherwise stated, we fix the central hard kinematics to
	\begin{equation}
		\begin{aligned}
			&Q^2=4~{\rm GeV}^2,
			\qquad
			x_g=10^{-3},
			\\
			&|\Pperp|=1.5~{\rm GeV},
			\qquad
			|\Dperp|=0.10~{\rm GeV}.
		\end{aligned}
		\label{eq:numerical-kinematics}
	\end{equation}
	The rapidity entering the dipole amplitude is fixed by
	$x_g=10^{-3}$,
	\begin{equation}
		Y
		=
		\ln\frac{1}{x_g}
		=
		\ln 10^3
		\simeq
		6.908.
		\label{eq:numerical-rapidity}
	\end{equation}
	Here $x_g$ is treated as an external small-$x$ kinematic parameter
	throughout the fixed-kinematics analysis.
	The Bessel--Gaussian envelope is set to
	\begin{equation}
		R_\gamma=30~{\rm GeV}^{-1}.
		\label{eq:numerical-Rgamma}
	\end{equation}
	The baseline moments are quoted in the fixed kernel convention of
	Eq.~\eqref{eq:BG-kernel}. Common electromagnetic, incident-flux, and
	final-state phase-space factors are omitted because the present
	benchmark concerns normalized responses rather than absolutely
	normalized cross sections. The quantities $\Sigma_{i,M}$ and
	$W_{i,M}$ are therefore reported in common arbitrary units. The quantities
	$\Sigma_{i,M}$ and $W_{i,M}$ are therefore reported in common
	arbitrary units. Their nonzero values, their dependence on $q_T$ and
	the beam--target geometry, and their role in the normalized
	combinations $v_{2,0}^{(M)}$, $D_M$, $C_M$, and $S_M$ are nevertheless
	well defined within the adopted convention.
	
	We choose the transverse recoil direction as the reference axis,
	\begin{equation}
		\phi_\Delta=0.
		\label{eq:phiDelta-convention}
	\end{equation}
	The fixed-offset results are evaluated in the collinear geometry
	\begin{equation}
		\Baxis
		=
		(B_{\rm ax},0),
		\qquad
		\phi_B-\phi_\Delta=0,
		\label{eq:collinear-offset-geometry}
	\end{equation}
	so that the beam--target displacement is parallel to the transverse
	recoil momentum. For the channel comparison and the reference
	$M=3$ calculation, we take
	\begin{equation}
		B_{\rm ax}
		=
		1~{\rm GeV}^{-1}.
		\label{eq:numerical-Baxis}
	\end{equation}
	We then scan the OAM channels
	\begin{equation}
		M=1,2,3,4,5,
		\label{eq:numerical-M-range}
	\end{equation}
	and vary the radial analyzer scale over
	\begin{equation}
		0.1\leq q_T\leq0.9~{\rm GeV}.
		\label{eq:numerical-qT-range}
	\end{equation}
	The broad channel scan contains 81 equally spaced values of $q_T$
	for each OAM channel,
	\begin{equation}
		q_T
		=
		0.10,\ 0.11,\ \ldots,\ 0.90~{\rm GeV},
		\label{eq:broad-qT-grid}
	\end{equation}
	corresponding to a spacing
	$\Delta q_T=0.01~{\rm GeV}$.
	
	The OAM kernel is evaluated at the light-front transverse coordinate
	defined in Eq.~\eqref{eq:kernel-coordinate-general}.
	The angular distributions are evaluated in the effective
	longitudinal channel defined in
	Eqs.~\eqref{eq:benchmark-dsigma0} and
	\eqref{eq:benchmark-dsigma1}, and the moments
	$\Sigma_{i,M}$ and $W_{i,M}$ are then constructed according to
	Eqs.~\eqref{eq:Sigma-iM-def} and \eqref{eq:W-iM-def}. Across the full numerical scan, the identity
	\begin{equation}
		S_M=D_M-C_M
		\label{eq:numerical-identity-check}
	\end{equation}
	is satisfied with a maximum absolute closure error below $3\times10^{-16}$.
	
	\subsection{OAM-channel dependence and node selectivity}
	\label{subsec:channel-dependence}
	
\begin{figure}[htbp]
		\centering
		\includegraphics[width=0.98\columnwidth]{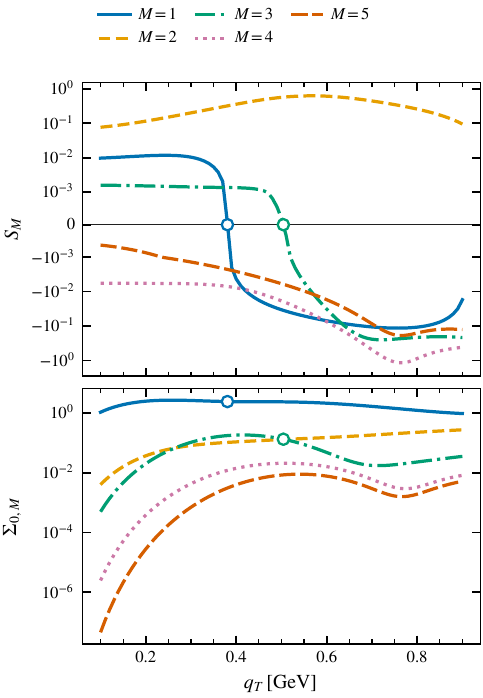}
		\caption{
			OAM-channel dependence of the normalized elliptic response and the
			baseline diffractive weight in the fixed kernel convention.
			The upper panel shows $S_M(q_T)$ for $M=1$--$5$ using a symmetric logarithmic scale.
			Open markers indicate the refined finite-yield zeros in the $M=1$ and $M=3$ channels.
			The lower panel shows the corresponding baseline weights
			$\Sigma_{0,M}(q_T)$ in the fixed analyzer convention on a logarithmic
			scale. Only selected OAM channels exhibit sign reversals. The relative
			weights decrease for the higher-$M$ profiles within the adopted
			kernel convention.
		}
		\label{fig:multiM-overview}
	\end{figure}

	Figure~\ref{fig:multiM-overview} summarizes the dependence of the
	normalized response and the baseline weight on the OAM channel. The upper panel shows $S_M(q_T)$ for $M=1$--$5$. Since the response spans both signs and approaches zero over part of the scan, a symmetric logarithmic scale is used to resolve small values without removing the sign information. The lower panel shows the corresponding baseline weights
	$\Sigma_{0,M}(q_T)$ in the fixed kernel convention of
	Eq.~\eqref{eq:BG-kernel}, plotted on a logarithmic scale.
	
	The scan reveals a pronounced channel selectivity. The $M=1$ and $M=3$ responses change sign within the interval in Eq.~\eqref{eq:numerical-qT-range}, whereas the $M=2$ response remains positive and the $M=4$ and $M=5$ responses remain negative. The occurrence of a node therefore does not follow a simple even--odd pattern. Instead, it reflects the channel-dependent radial overlap between the Bessel--Gaussian analyzer and the impact-parameter-dependent dipole amplitude.
	
	The refined $M=1$ zero is located at
	\begin{equation}
		q_{T,1}^{\ast}
		\simeq
		0.3806~{\rm GeV},
		\label{eq:qstar-M1}
	\end{equation}
	with a finite baseline weight in the adopted kernel convention,
	\begin{equation}
		\Sigma_{0,1}
		\left(
		q_{T,1}^{\ast}
		\right)
		\simeq
		2.43.
		\label{eq:sigma0-M1}
	\end{equation}
	At this point,
	\begin{equation}
		D_1
		\simeq
		C_1
		\simeq
		-3.2\times10^{-7},
		\label{eq:M1-DC-node}
	\end{equation}
	so the equality $D_1=C_1$ is realized close to a simultaneous zero of both contributions. A local decomposition of the $q_T$ dependence shows that the sign reversal is controlled predominantly by the direct term $D_1$, while the variation of $C_1$ is much smaller. The $M=1$ crossing therefore realizes approximately the direct-response mechanism defined in Eq.~\eqref{eq:direct-response-node}.
	
	The $M=3$ channel exhibits a qualitatively different node. Although its baseline weight is smaller than that of the $M=1$
	channel within the adopted kernel convention, the crossing occurs through the cancellation of two individually finite contributions. We therefore use $M=3$ as the principal example of a nontrivial finite-yield OAM projection node.
	
	The channel comparison also illustrates why the normalized response
	should be displayed together with the corresponding baseline weight.
	A large value of $|S_M|$ does not by itself imply a comparably large
	contribution in the adopted wave-packet convention. In particular,
	the higher-$M$ profiles can develop sizable normalized responses in
	regions where the calculated $\Sigma_{0,M}$ is small. Since an
	absolute comparison of experimental event rates would additionally
	require a specified incident-flux normalization for each prepared
	OAM mode, the lower panel is used here as a relative numerical
	diagnostic rather than as a direct rate prediction.
	
	\subsection{Nontrivial finite-yield node in the \(M=3\) channel}
	\label{subsec:finite-yield-zero-results}
	
\begin{figure}[htbp]
		\centering
		\includegraphics[width=0.98\columnwidth]{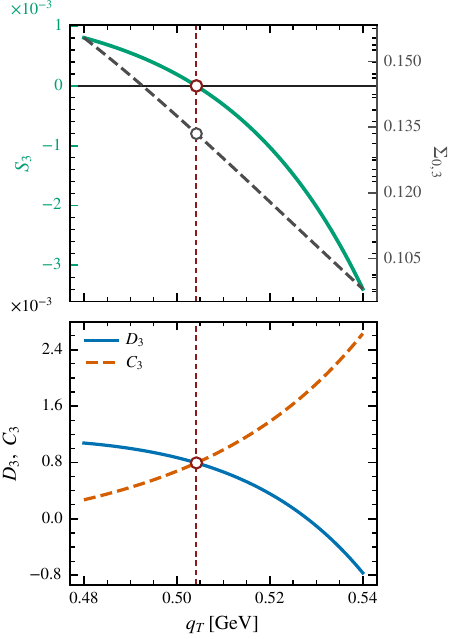}
		\caption{
			Nontrivial finite-yield projection node in the $M=3$ channel for $B_{\rm ax}=1~{\rm GeV}^{-1}$.
			The upper panel shows the normalized response $S_3(q_T)$ on the left axis and the baseline weight $\Sigma_{0,3}(q_T)$ on the right axis.
			The response changes sign at
			$q_{T,3}^{\ast}\simeq0.5042~{\rm GeV}$
			while the baseline weight remains finite in the fixed kernel
			convention.
			The lower panel shows the direct contribution $D_3$ and the normalization contribution $C_3$.
			Their finite-valued crossing generates the node.
		}
		\label{fig:response-zero}
	\end{figure}

	Figure~\ref{fig:response-zero} presents the refined $M=3$ result. The upper panel displays the normalized response $S_3(q_T)$ on the
	left axis and the baseline weight $\Sigma_{0,3}(q_T)$ on the right
	axis. The response changes sign at
	\begin{equation}
		q_{T,3}^{\ast}
		\simeq
		0.5042~{\rm GeV},
		\label{eq:qstar-main}
	\end{equation}
	while the baseline weight remains finite in the same kernel
	convention,
	\begin{equation}
		\Sigma_{0,3}
		\left(
		q_{T,3}^{\ast}
		\right)
		\simeq
		0.134.
		\label{eq:sigma0-star-main}
	\end{equation}
	Its numerical value is convention dependent, but its nonvanishing
	character and the normalized cancellation condition are not.
	
	The lower panel of Fig.~\ref{fig:response-zero} resolves the mechanism through
	\begin{equation}
		S_3(q_T)
		=
		D_3(q_T)-C_3(q_T).
		\label{eq:DC-decomposition-main}
	\end{equation}
	At the refined node,
	\begin{equation}
		D_3
		\simeq
		C_3
		\simeq
		7.94\times10^{-4}.
		\label{eq:M3-DC-values}
	\end{equation}
	The two terms are therefore individually finite. On the low-$q_T$ side of the node,
	\begin{equation}
		D_3>C_3,
	\end{equation}
	and hence $S_3>0$. As $q_T$ increases, $D_3$ decreases while $C_3$ increases. Their crossing reverses the inequality,
	\begin{equation}
		D_3<C_3,
	\end{equation}
	and produces $S_3<0$.
	
	The $M=3$ node consequently realizes the compensation mechanism in Eq.~\eqref{eq:compensation-node}. It is not a zero of the direct response, the normalization response,
	the dipole amplitude, or the baseline diffractive weight. It appears only after the direct and normalization contributions are combined into the normalized observable.
	
	At the node, the leading deformation-induced variation of the elliptic moment is suppressed,
	\begin{equation}
		A_2^{(3)}(\epsilon_2)
		-
		A_2^{(3)}(0)
		=
		\mathcal O(\epsilon_2^2),
		\label{eq:A2-suppression-main}
	\end{equation}
	because the coefficient of the linear term vanishes. The baseline moment $v_{2,0}^{(3)}$ and the underlying scattering strength may nevertheless remain nonzero. The node therefore characterizes the sensitivity of the normalized readout to the target quadrupole deformation, rather than the existence of the elliptic target component itself.
	
	\subsection{Local response density across the node}
	\label{subsec:local-density-results}
	
	To identify the transverse origin of the cancellation, we construct the signed local contribution to the normalized response. For a transverse bin centered at $\Xperp$, we define
	\begin{equation}
		\dd S_3(\Xperp)
		=
		\frac{
			\dd W_{1,3}(\Xperp)
			-
			v_{2,0}^{(3)}
			\dd\Sigma_{1,3}(\Xperp)
		}{
			\Sigma_{0,3}
		}.
		\label{eq:local-dS-def}
	\end{equation}
	The corresponding density satisfies
	\begin{equation}
		\int
		\dd^2\Xperp\,
		\frac{\dd S_3}{\dd^2\Xperp}
		=
		S_3.
		\label{eq:local-dS-closure}
	\end{equation}
	This quantity is a signed response density and not a probability density. Positive and negative regions represent local contributions whose integral gives the global normalized response.
	
\begin{figure*}[htbp]
		\centering
		\includegraphics[width=0.98\textwidth]{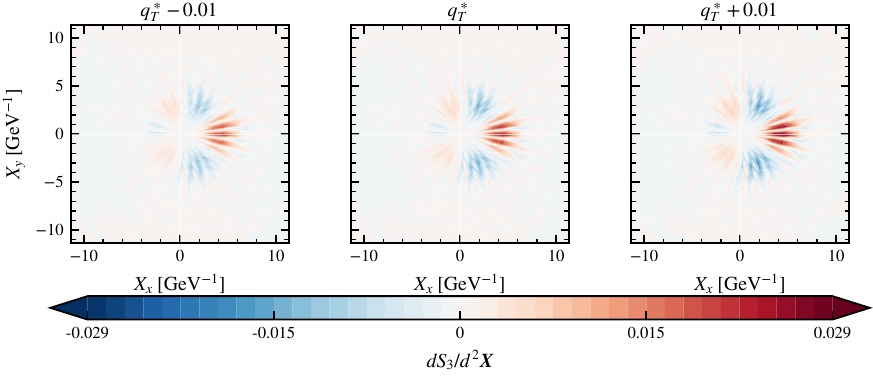}
		\caption{
			Signed local response density $\dd S_3/\dd^2\Xperp$ across the $M=3$ finite-yield projection node.
			The panels correspond approximately to $q_T=q_{T,3}^{\ast}-0.01~{\rm GeV}$, $q_T=q_{T,3}^{\ast}$, and $q_T=q_{T,3}^{\ast}+0.01~{\rm GeV}$.
			A common symmetric color scale is used.
			The global response changes from positive to nearly zero and then to negative through a redistribution of signed transverse contributions.
		}
		\label{fig:density-triptych}
	\end{figure*}

	Figure~\ref{fig:density-triptych} shows the local response density below, near, and above the $M=3$ projection node. The three panels are evaluated at
	$
	q_T
	\simeq
	q_{T,3}^{\ast}-0.01~{\rm GeV},\ 
	q_{T,3}^{\ast},\ 
	q_{T,3}^{\ast}+0.01~{\rm GeV}
	$
	and use a common symmetric color scale.
	
	Below the node, the positive local contributions dominate after integration, producing $S_3>0$. Near the node, substantial positive and negative domains remain present, but their integrated contributions nearly compensate. Above the node, the negative contribution becomes dominant and the global response changes sign. The projection node is therefore a global cancellation of spatially resolved response contributions, rather than a pointwise disappearance of the OAM-weighted scattering structure.
	
	For the three density maps, the integrated local response reproduces
	the independently evaluated global response with a maximum closure
	error below $2\times10^{-17}$. This provides a direct consistency
	check of the local decomposition and confirms that the sign reversal
	is represented by the redistribution of the signed transverse
	response.
	
	\subsection{Collinear offset dependence}
	\label{subsec:alignment-dependence}
	
	We next examine the dependence of the projection node on the
	magnitude of the beam--target offset within the collinear geometry
	defined in Eq.~\eqref{eq:collinear-offset-geometry}. Increasing $B_{\rm ax}$ displaces the target along the recoil
	direction relative to the radial and phase structure of the
	Bessel--Gaussian analyzer . This modifies both $D_3$ and $C_3$, and therefore changes the solution of the cancellation condition.
	
\begin{figure}[htbp]
		\centering
		\includegraphics[width=0.98\columnwidth]{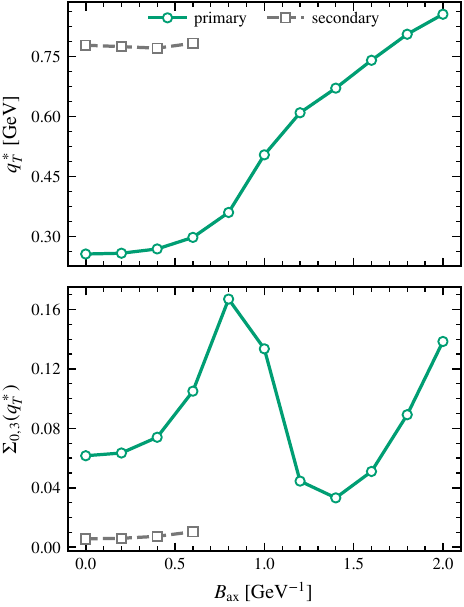}
		\caption{
			Collinear offset dependence of the $M=3$ finite-yield projection
			nodes.
			The upper panel shows the refined node positions as functions of
			$B_{\rm ax}$ for $\phi_B-\phi_\Delta=0$.
			The solid line denotes the primary, lowest-$q_T$ branch.
			A secondary higher-$q_T$ branch appears at small offsets and is shown by the dashed line.
			The lower panel gives the corresponding baseline weights
			$\Sigma_{0,3}(q_T^\ast)$ in the fixed kernel convention.
		}
		\label{fig:alignment}
	\end{figure}

	Figure~\ref{fig:alignment} shows the refined node positions as a function of the fixed offset magnitude $B_{\rm ax}$. The solid curve identifies the lowest-$q_T$ finite-yield solution and is referred to as the primary branch. As the offset varies over
	\begin{equation}
		0
		\leq
		B_{\rm ax}
		\leq
		2~{\rm GeV}^{-1},
		\label{eq:Bax-scan-range}
	\end{equation}
	the primary node moves from approximately
	\begin{equation}
		q_T^\ast
		\simeq
		0.256~{\rm GeV}
	\end{equation}
	to
	\begin{equation}
		q_T^\ast
		\simeq
		0.856~{\rm GeV}.
	\end{equation}
	The accompanying baseline weight remains finite throughout the
	displayed branch.
	
	The baseline weight along the primary branch is nonmonotonic. It decreases over an intermediate range of offsets and rises again at larger $B_{\rm ax}$. This behavior reflects the structured overlap between the displaced target profile and the radial Bessel weighting; it cannot be reduced to a simple monotonic loss of transverse overlap.
	
	For small offsets, a second higher-$q_T$ crossing is also found in the scanned interval. This secondary branch occurs around
	\begin{equation}
		q_T^\ast
		\simeq
		0.77\text{--}0.78~{\rm GeV}
	\end{equation}
	for $B_{\rm ax}\lesssim0.6~{\rm GeV}^{-1}$. Its baseline weight is only a few percent of the maximal weight at the
	corresponding offset within the same kernel convention. We therefore distinguish it from the primary finite-yield branch and display it as a secondary solution. Determining whether this branch moves beyond the present $q_T$ range, merges with another solution, or terminates as the offset increases would require an extended high-$q_T$ scan.
	
	The strong migration of the primary node demonstrates that the cancellation is not a universal zero of the target amplitude. It is controlled by the relative placement of the localized OAM analyzer and the target. In an experimental setting, a measured observable would involve a distribution of beam--target offsets, finite overlap, or an equivalent event selection. The fixed-offset scan isolates the underlying geometric mechanism and shows that the OAM wave packet supplies a tunable transverse projection of the same small-$x$ target matrix element.
	
	The numerical values in this section are specific to the selected dipole profile, Bessel--Gaussian analyzer, and benchmark kinematics. They should therefore be interpreted as evidence for the observable mechanism and its channel dependence, rather than as direct experimental predictions. A quantitative phenomenological analysis would require realistic dipole amplitudes constrained by diffractive data, small-$x$ evolution, finite-offset averaging, detector acceptance, and event-level jet reconstruction.

	\section{Conclusions}
	\label{sec:conclusions}
	
	We have developed a localized orbital-angular-momentum-resolved
	extension of elliptic small-$x$ gluon tomography in hard diffractive
	dijet deep inelastic scattering. The OAM dependence is introduced
	through the transverse structure of the localized lepton-side
	analyzer, while the target remains described by the same
	impact-parameter-dependent dipole matrix element that enters
	plane-wave diffractive tomography
	~\cite{Hatta:2016dxp,Zhou:2016rnt,Hagiwara:2017fye,
		Mantysaari:2019csc,Hatta:2024ocp,Shao:2024gri}.
	The construction therefore changes how the target geometry is
	projected into the observable rather than modifying the underlying
	small-$x$ target distribution.
	
	The central quantity is the normalized linear response
	$S_M=D_M-C_M$ of the OAM-resolved elliptic moment to the target
	quadrupole deformation. Here $D_M$ is the direct response of the
	elliptic numerator, whereas $C_M$ arises from the simultaneous
	deformation of the normalization. A finite-yield projection node is
	defined by $S_M=0$ with $\Sigma_{0,M}\neq0$. The $M=1$ crossing is
	approximately associated with a zero of the direct response, while
	the $M=3$ channel exhibits the more distinctive compensation
	mechanism $D_3=C_3\neq0$. For the reference collinear configuration,
	the latter occurs at
	$q_{T,3}^{\ast}\simeq0.504~{\rm GeV}$ with a nonvanishing baseline
	weight in the fixed analyzer convention. The node condition and the
	decomposition into $D_M$ and $C_M$ are invariant under an overall
	rescaling of the transverse kernel.
	
	The signed local response density shows that the $M=3$ node is a
	global cancellation of spatially resolved positive and negative
	contributions rather than a disappearance of the underlying
	OAM-weighted scattering structure. Moreover, varying the magnitude
	of the collinear beam--target offset shifts the node continuously
	over a broad range of $q_T$, demonstrating that the cancellation is
	controlled by the transverse overlap between the localized analyzer
	and the target. This behavior is consistent with the general role of
	transverse localization and impact-parameter geometry in
	phase-sensitive vortex scattering
	~\cite{Ivanov:2011bv,Ivanov:2011aa,Ivanov:2012na,
		Karlovets:2015nva,Karlovets:2016jrd,Karlovets:2016uhb,
		Yang:2026byv}.
	
	The present calculation is intended as a proof-of-principle study of
	this projection mechanism rather than as an absolute phenomenological
	prediction. Quantitative applications will require dipole amplitudes
	constrained by diffractive data, realistic small-$x$ evolution,
	finite distributions of beam--target offsets, a more complete
	treatment of the lepton and photon channels, and experimental
	acceptance and dijet reconstruction
	~\cite{GolecBiernat:1998js,Kowalski:2003hm,Kowalski:2006hc,
		Rezaeian:2013tka,Salazar:2019ncp,Iancu:2021rup,Hatta:2022lzj,
		Accardi:2012qut,AbdulKhalek:2021gbh}.
	The results nevertheless demonstrate that localized OAM wave packets
	provide a tunable transverse projection of elliptic small-$x$ gluon
	geometry, with the OAM channel, radial mode scale, and transverse
	offset controlling distinct aspects of the readout.

	% ============================================================
	% Acknowledgments
	% ============================================================
	\begin{acknowledgments}
		We thank Igor P. Ivanov for carefully reading an earlier version of
		the manuscript and for valuable comments and suggestions.
		This work has been supported by the National Natural Science Foundation of China
		(Grant No. 12547118) and the Research Program of State Key Laboratory of Heavy Ion
		Science and Technology, Institute of Modern Physics, Chinese Academy of Sciences
		(Grant No. HIST2025CS08).
	\end{acknowledgments}
	% ============================================================
	% Bibliography
	% ============================================================
	\bibliographystyle{apsrev4-2}
	\bibliography{refs}
	
\end{document}